\documentclass[manuscript,screen,nonacm]{acmart}

\setcopyright{none}
\usepackage{todonotes}
\usepackage{tcolorbox}
\tcbuselibrary{skins,breakable}
\usepackage{enumitem}
\usepackage{xcolor}
\usepackage{soul}
\usepackage{adjustbox}
\usepackage{booktabs}

\definecolor{Accent}{HTML}{2D7FF9}  
\definecolor{Panel}{HTML}{F8F9FB}   
\definecolor{Frame}{HTML}{E4E7EB}   
\definecolor{Ink}{HTML}{111111}     

\newtcolorbox{policybox}[2][]{%
  enhanced,
  breakable,
  colback=Panel,
  colframe=Frame,
  coltitle=Ink,
  title={\sffamily\bfseries #2},
  fonttitle=\sffamily\bfseries,
  arc=2mm,
  boxrule=0.4pt,
  left=2mm,right=2mm,top=1.5mm,bottom=1.5mm,
  boxsep=1mm,
  borderline west={2pt}{0pt}{Accent},
  before skip=6pt, after skip=6pt,
  sharp corners,
  drop shadow={black!10!white},
  #1
}

\AtBeginDocument{%
  }

\setcopyright{acmlicensed}
\copyrightyear{2026}
\acmYear{2026}
\acmDOI{XXXXXXX.XXXXXXX}

\acmJournal{AILET}
\acmVolume{37}
\acmNumber{4}
\acmArticle{111}
\acmMonth{8}

\begin{document}

\title{Identity or Prompt Noise? A Calibrated Invariance Audit of LLM Code Generation}

\author{Maksim E. Eren}
\email{maksim@lanl.gov}
\orcid{0000-0002-4362-0256}
\affiliation{%
  \institution{Computational Intelligence \& Modeling, LANL}
  \city{Los Alamos}
  \state{NM}
  \country{USA}
}

\author{Ryan Barron}
\email{barron@lanl.gov}
\orcid{0000-0002-4362-0256}
\affiliation{%
  \institution{Computational Intelligence \& Modeling, LANL}
  \city{Los Alamos}
  \state{NM}
  \country{USA}
}

\author{Eric Michalak}
\email{emichalak@lanl.gov}
\orcid{0000-0002-4362-0256}
\affiliation{%
  \institution{Advanced Research in Cyber Systems, LANL}
  \city{Los Alamos}
  \state{NM}
  \country{USA}
}

\author{Charles Nicholas}
\email{nicholas@umbc.edu}
\orcid{0000-0002-4362-0256}
\affiliation{%
  \institution{Computer Science and Electrical Engineering, UMBC}
  \city{Baltimore}
  \state{MD}
  \country{USA}
}

\author{Manish Bhattarai}
\email{ceodspspectrum@lanl.gov}
\orcid{0000-0002-4362-0256}
\affiliation{%
  \institution{Theoretical Division, LANL}
  \city{Los Alamos}
  \state{NM}
  \country{USA}
}


\begin{abstract}
Identity cues are irrelevant to a fixed programming specification, but raw counterfactual differences can arise from unequal samples and prompt wording. We audit 30.73 million executed Python generations from seven checkpoints on HumanEval+ and MBPP+, supplemented by an exploratory 550B slice, under model-assigned gender, country, and occupation personas. Within-task randomization and false-discovery-rate control identify occupation as the most consistent structural axis: CodeBLEU dispersion exceeds its exchangeability null in 10/14 model--benchmark cells, remains significant in 8/12 full-coverage cells, and exceeds the country ratio in every paired cell, although the median excess is only 0.141 points. In six high-pass-rate cells, occupation dispersion replicates across token similarity, length, comments, reference similarity, and complexity, while pass-rate dispersion is significant in none. Country leads raw dispersion in 12/14 cells but has a median calibrated ratio of 1.00. Gender-associated variation cannot be separated from persona wording in this design. Thus, the evidence supports small, reproducible occupation-conditioned changes in code form, not stable disadvantage to named identities or demonstrated downstream harm. We use this finding to motivate a broader commentary on the potential impacts of bias in code generation, including the possibility that models may condition their outputs on identity information available from prior conversational context. 
\end{abstract}




\maketitle

\section{Introduction}
\label{sec:introduction}
Functional correctness does not establish that a code generator is invariant to task-irrelevant context. For a fixed programming task, changing a model-assigned gender, country, or occupation persona should alter neither correctness nor implementation form. Yet unequal samples across identity values and innocuous rewording can create raw differences, so an identity effect cannot be inferred from extreme groups or neutral-relative deltas alone. Prior work examines demographic stereotypes in language models~\cite{gallegos2024bias,sheng2019woman}, discriminatory logic in socially sensitive programs~\cite{liu2023uncovering,huang2025bias}, and code sensitivity to personas or prompt reformulation~\cite{guo2025personality,paleyes2026code}. We instead test \emph{irrelevant-identity invariance}: executable tasks remain fixed while the assigned identity persona varies. This separates harmful application logic from an upstream robustness question: whether task-irrelevant identity cues alter outputs for an unchanged programming specification. We therefore evaluate functional correctness and code form separately.

Our contribution is a calibrated audit of 30.73 million executed generations. A within-task randomization test asks whether identity-conditioned dispersion exceeds a label-exchangeability null; false-discovery-rate (FDR) correction, wording controls, a full-coverage cohort, and matched-cell scale comparisons bound the result. Calibration reverses the raw conclusion: country appears most variable before calibration, whereas occupation is the most reproducible structural axis afterward. The effect is small and concerns code form, not demonstrated harm or stable treatment of named occupations.

\section{Experimental Setup and Results}
\label{sec:exp_results}
\subsection{Design and Estimand}

We test counterfactual invariance: holding task and generation settings fixed, does varying a task-irrelevant assigned persona alter correctness or implementation form? Our estimand is across-identity dispersion in task-centered outcomes beyond that induced by sample allocation; it tests axis-level non-exchangeability, not benefit, harm, or causality for named identities.

HumanEval+ (164 tasks) and MBPP+ (378 tasks) provide augmented functional tests~\cite{chen2021humaneval,austin2021mbpp,liu2023evalplus}. Fixed tasks are paired with 20 gender labels, 112 countries~\cite{haerpfer2022wvs}, 234 occupations, and ten generic persona-noun controls. Every exact prompt is decoded once; identities recur across tasks. The main archive contains 30{,}728{,}903 generations from seven checkpoints (Llama-3.2-3B, Mistral-7B, Llama-3-8B, Nemotron-3-Nano-30B, Gemma-4-31B, gpt-oss-120B, and Nemotron-3-Super-120B)~\cite{dubey2024llama3,jiang2023mistral,nvidia2025nemotron3nano30b,gemmateam2026gemma4,openai2025gptoss,nvidia2026nemotron3super120b}. Gemma-4-31B has incomplete coverage; an additional 260{,}000-row Ultra-550B HumanEval+ occupation slice is exploratory. We infer within model, benchmark, and axis, repeat the analysis on six full-coverage models, and match cells for scale comparisons.

Generation settings are fixed within runs (temperature 0.2, top-$p$ 0.95, and at most 2{,}048 new tokens). Parseable programs are sandboxed and checked against EvalPlus tests. EvalPlus pass rate is the functional primary outcome. CodeBLEU, which combines lexical, syntactic, and data-flow similarity~\cite{ren2020codebleu}, is the structural primary outcome; it measures similarity to a reference, not code quality or benefit. Five additional structure metrics are exploratory.

For each model--benchmark--axis cell, we center outcomes within task and compute the standard deviation $\widehat{\sigma}_{\mathrm{obs}}$ of identity-level means. We then shuffle outcomes within task 1{,}000 times, preserving tasks, sample sizes, and outcome distributions while breaking identity association. We report the observed-to-null ratio, one-sided randomization $p$-value, and excess dispersion
\[
\widehat{\sigma}_{\mathrm{excess}}=\sqrt{\max(0,\widehat{\sigma}_{\mathrm{obs}}^2-\widehat{\sigma}_{\mathrm{null}}^2)}.
\]
This is a label-exchangeability null, not an exact-prompt decoding baseline. Benjamini--Hochberg correction at $q<0.05$ is applied separately to 42 CodeBLEU and 42 pass-rate tests (seven models, two benchmarks, three axes), then recomputed over 36 tests for the full-coverage cohort. Neutral persona-noun and within-gender noun controls form separate corrected families.

\subsection{Results}

Table~\ref{tab:calibration-summary} gives the confirmatory result. Occupation CodeBLEU dispersion survives correction in 10/14 main cells and 8/12 full-coverage cells; its ratio exceeds one in all 14 and exceeds country in every paired cell. Country nevertheless has the largest \emph{raw} spread in 12/14 cells because it has fewer observations per value. After calibration its median ratio is 1.00. Thus calibration changes which axis appears dominant, not merely its reported magnitude. Functional discoveries are less consistent, and median pass-rate excess is at most 0.315 percentage points. Given the corpus size, we interpret significance together with absolute excess.

\begin{table}[t]
\centering
\small
\setlength{\tabcolsep}{4pt}
\caption{Calibrated dispersion. FDR counts BH-significant cells; ratios and CodeBLEU excess are main-cohort medians. ``Raw lead'' counts the largest uncalibrated CodeBLEU spread.}
\label{tab:calibration-summary}
\begin{adjustbox}{max width=\linewidth}
\begin{tabular}{lrrrrrr}
\toprule
& \multicolumn{3}{c}{\textbf{CodeBLEU}} & \multicolumn{2}{c}{\textbf{Pass}} & \\
\cmidrule(lr){2-4}\cmidrule(lr){5-6}
\textbf{Axis} & \textbf{Main FDR} & \textbf{Full FDR} & \textbf{Ratio/Excess} & \textbf{Main FDR} & \textbf{Full FDR} & \textbf{Raw lead} \\
\midrule
Occupation & 10/14 & 8/12 & 1.29 / 0.141 & 6/14 & 4/12 & 1/14 \\
Gender     &  7/14 & 7/12 & 1.35 / 0.095 & 6/14 & 4/12 & 1/14 \\
Country    &  2/14 & 1/12 & 1.00 / 0.021 & 1/14 & 1/12 & 12/14 \\
\bottomrule
\end{tabular}
\end{adjustbox}
\end{table}

\begin{figure}[t]
  \centering
  \includegraphics[width=0.94\linewidth]{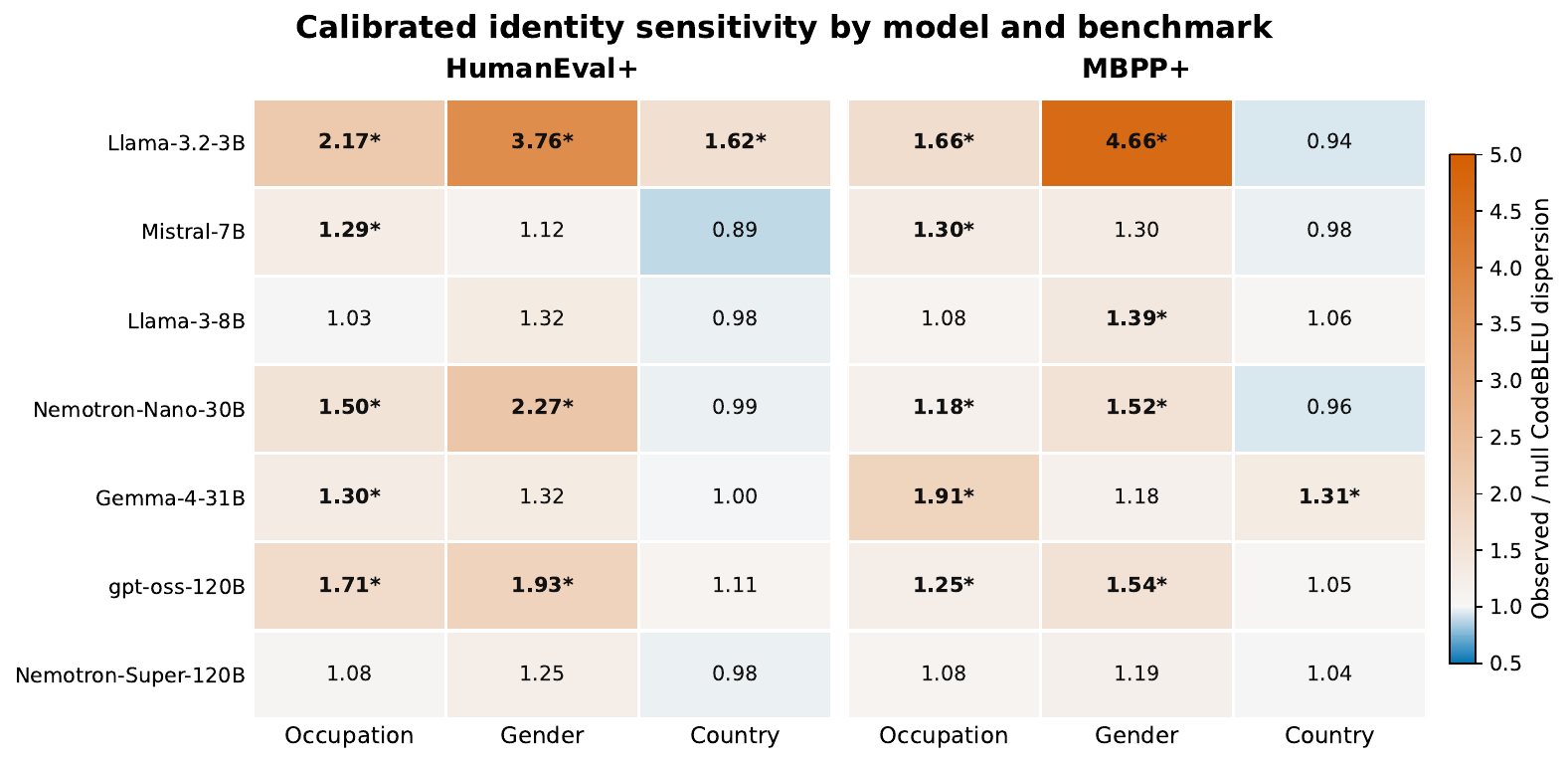}
  \caption{Observed-to-null CodeBLEU dispersion across identity axes. Asterisks denote BH-adjusted $q<0.05$; Gemma-4-31B results reflect incomplete coverage.}
  \label{fig:calibrated-ratios}
\end{figure}

Figure~\ref{fig:calibrated-ratios} shows every main-checkpoint CodeBLEU cell. Occupation replicates on both benchmarks for five of seven checkpoints, compared with three for gender and none for country. The two country discoveries are isolated rather than cross-benchmark effects.

Wording controls constrain the gender interpretation. Neutral persona nouns are significant in 2/14 CodeBLEU cells and 0/14 pass cells. Regrouping gender-only rows by persona noun yields 6/14 CodeBLEU and 4/14 pass discoveries, similar to gender's 7/14 and 6/14; paired ratios do not differ ($p=0.358$, exploratory). Thus demographic meaning is not isolated from wording.

Five exploratory structure metrics are corrected jointly across nine available model--benchmark cells (45 tests). Table~\ref{tab:triangulation} summarizes the six complete cells: token BLEU, lines, and comments are significant in all six, and reference similarity and complexity in five; pass rate, corrected in its separate primary family, is significant in none. The effects are consistent but small.

\begin{table}[t]
\centering
\small
\setlength{\tabcolsep}{6pt}
\caption{Occupation sensitivity in six full, high-pass-rate cells.}
\label{tab:triangulation}
\begin{tabular}{lrrr}
\toprule
\textbf{Metric} & \textbf{FDR cells} & \textbf{Median ratio} & \textbf{Median excess} \\
\midrule
Reference similarity & 5/6 & 1.916 & 0.578 points \\
Token BLEU            & 6/6 & 1.620 & 0.403 points \\
Lines of code         & 6/6 & 1.823 & 0.189 lines \\
Comment lines         & 6/6 & 1.369 & 0.021 lines \\
Complexity            & 5/6 & 1.108 & 0.011 units \\
EvalPlus pass         & 0/6 & 1.041 & 0.140 points \\
\bottomrule
\end{tabular}
\end{table}

On 15{,}312 cells shared by all eight archived checkpoints, parameter count correlates with pass rate (Spearman $\rho=0.922$, $p=0.001$) but not with lower occupation spread ($\rho=0.431$, $p=0.286$). The exploratory Ultra-550B slice likewise has calibrated occupation CodeBLEU sensitivity (ratio 1.305, $p=0.001$) but null pass sensitivity (ratio 1.020, $p=0.349$). However, occupation rankings are unstable across checkpoints (median pairwise Spearman $\rho=0.067$). Scale therefore does not guarantee invariance, but the data do not support stable favoritism toward named occupations or a causal scaling law.

\section{Discussion and Limitations}
\label{sec:commentary}
The defensible finding is narrow: task-irrelevant occupation personas predict small changes in implementation form across models, benchmarks, and independent metrics, while correctness is more stable. This is \emph{bias} only as identity-conditioned non-invariance. Reference similarity is not inherently beneficial, personas are model assignments rather than observed users, and we do not measure security, validated maintainability, developer effort, or social harm.

This distinction matters for evaluation. An accuracy-only audit misses structural sensitivity, while an uncalibrated audit selects country as the dominant axis in 12/14 cells even though its calibrated ratio is nearly null. The proposed test supplies a release-time statistic on fixed executable tasks and distinguishes robust axis-level dispersion from noisy identity extremes; the same calibration applies to other irrelevant contextual cues. The near-equivalence of gender and wording discoveries also exposes a confound that neutral-baseline comparisons alone do not resolve.

The archive permits two downstream-adjacent checks. Maintainability proxies show reproducible but minute occupation effects (median excess: 0.189 lines, 0.021 comment lines, and 0.011 complexity units), not validated maintainability outcomes. In a separate exploratory family of 30 tests over the same six high-pass cells, failures under augmented tests and timeouts are significant in 0/6 cells each, harness execution time in 1/6, and execution failures and imports in 3/6 each. Corresponding median excesses are only 2.98 ms, 0.134 percentage points, and 0.002 imports; wall time is not controlled algorithmic efficiency, and imports are not security findings.

Occupation pass dispersion is significant in 6/14 main cells, but all six discoveries occur in Mistral-7B, Llama-3-8B, or Gemma-4-31B, and none occurs in the six full high-pass cells. Across 234 occupations and 12 full model--benchmark cells, task-normalized pass rankings have median pairwise Spearman $\rho=0.001$ (range $-0.231$ to $0.462$), and no occupation remains in the bottom quartile in at least 9/12 cells. This argues against stable correctness disadvantage in the present benchmarks, not against discrimination generally.

Three limitations bound interpretation. First, the permutation test estimates label exchangeability within task, not stochastic variation from repeated decoding of the exact prompt. Existing persona nouns are only partially crossed with occupations; a retrospective held-out-task decomposition cannot establish a semantic identity effect because wording-only prediction is near zero and occupation-only prediction is negative. A causal test requires identities fully crossed with paraphrases and repeated decodes. Second, Gemma-4-31B and Ultra-550B have incomplete coverage; central results are also reported for a full-coverage cohort, and Ultra-550B supports only an exploratory occupation claim. Third, the five-metric and consequence-proxy analyses were conducted after the primary CodeBLEU finding and remain exploratory despite conservative correction. Coverage imbalance, model-family confounding, small absolute effects, and unstable occupation rankings further preclude claims about named groups or practical harm.

In summary, these results ask three simple questions: do identity labels change the generated code more than we would expect from sampling alone, does that pattern repeat across models and benchmarks, and is the difference large enough to matter? Calibration helps separate apparent differences from reproducible ones. After calibration, occupation shows a small but reproducible difference in how code is written, while most country differences largely disappear and gender effects cannot yet be separated from persona wording. Importantly, this should not be read as evidence that particular occupations are consistently favored or harmed: the affected occupations are not stable across models, and the differences do not consistently change whether the code passes its tests. Existing proxies show no consistent downstream penalty, and the present results do not establish effects in conversational or educational settings. Still, even small identity-conditioned differences in code form could matter if they affect properties not measured here, such as maintainability, security, or behavior over longer interactions. This may be especially relevant in conversational systems, where identity information disclosed or inferred earlier in a chat---or retained through conversational memory---could become part of the context used for later code-generation requests even when that information is irrelevant to the programming task. We therefore use these findings to motivate a broader commentary on potential robustness risks, including possible implications for education, where such variation could influence the examples, explanations, or programming support different learners receive. We do not interpret the present results as evidence of demonstrated downstream harm. Future work should test repeated exact prompts, fully crossed paraphrases, longer conversational contexts, and downstream outcomes in settings such as software security, maintenance, and education to determine when these differences become consequential.

\section*{Acknowledgments}
Approved for unlimited release LA-UR-26-27647. This work was funded by Los Alamos National Laboratory, operated by Triad National Security, LLC, for the National Nuclear Security Administration of the U.S. Department of Energy (Contract No. 89233218CNA000001).

\bibliographystyle{ACM-Reference-Format}
\bibliography{references}

@article{gallegos2024bias,
  title     = {Bias and Fairness in Large Language Models: A Survey},
  author    = {Gallegos, Isabel O. and
               Rossi, Ryan A. and
               Barrow, Joe and
               Tanjim, Md Mehrab and
               Kim, Sungchul and
               Dernoncourt, Franck and
               Yu, Tong and
               Zhang, Ruiyi and
               Ahmed, Nesreen K.},
  journal   = {Computational Linguistics},
  volume    = {50},
  number    = {3},
  pages     = {1097--1179},
  year      = {2024},
  month     = sep,
  address   = {Cambridge, MA},
  publisher = {MIT Press},
  doi       = {10.1162/coli_a_00524},
  url       = {https://aclanthology.org/2024.cl-3.8/}
}

@inproceedings{sheng2019woman,
  title     = {The Woman Worked as a Babysitter: On Biases in Language Generation},
  author    = {Sheng, Emily and
               Chang, Kai-Wei and
               Natarajan, Premkumar and
               Peng, Nanyun},
  booktitle = {Proceedings of the 2019 Conference on Empirical Methods
               in Natural Language Processing and the 9th International
               Joint Conference on Natural Language Processing
               (EMNLP-IJCNLP)},
  pages     = {3407--3412},
  year      = {2019},
  month     = nov,
  address   = {Hong Kong, China},
  publisher = {Association for Computational Linguistics},
  doi       = {10.18653/v1/D19-1339},
  url       = {https://aclanthology.org/D19-1339/}
}

@inproceedings{liu2023uncovering,
  title     = {Uncovering and Quantifying Social Biases in Code Generation},
  author    = {Liu, Yan and
               Chen, Xiaokang and
               Gao, Yan and
               Su, Zhe and
               Zhang, Fengji and
               Zan, Daoguang and
               Lou, Jian-Guang and
               Chen, Pin-Yu and
               Ho, Tsung-Yi},
  booktitle = {Advances in Neural Information Processing Systems},
  volume    = {36},
  pages     = {2368--2380},
  year      = {2023},
  publisher = {Curran Associates, Inc.},
  url       = {https://papers.nips.cc/paper_files/paper/2023/hash/071a637d41ea290ac4360818a8323f33-Abstract-Conference.html}
}

@article{huang2025bias,
  title     = {Bias Testing and Mitigation in {LLM}-based Code Generation},
  author    = {Huang, Dong and
               Zhang, Jie M. and
               Bu, Qingwen and
               Xie, Xiaofei and
               Chen, Junjie and
               Cui, Heming},
  journal   = {ACM Transactions on Software Engineering and Methodology},
  volume    = {35},
  number    = {1},
  pages     = {5:1--5:31},
  articleno = {5},
  numpages  = {31},
  year      = {2026},
  month     = jan,
  publisher = {Association for Computing Machinery},
  address   = {New York, NY, USA},
  doi       = {10.1145/3724117},
  url       = {https://doi.org/10.1145/3724117}
}

@inproceedings{guo2025personality,
  title     = {Personality-Guided Code Generation Using Large Language Models},
  author    = {Guo, Yaoqi and
               Chen, Zhenpeng and
               Zhang, Jie M. and
               Liu, Yang and
               Ma, Yun},
  booktitle = {Proceedings of the 63rd Annual Meeting of the Association
               for Computational Linguistics
               (Volume 1: Long Papers)},
  pages     = {1068--1080},
  year      = {2025},
  month     = jul,
  address   = {Vienna, Austria},
  publisher = {Association for Computational Linguistics},
  isbn      = {979-8-89176-251-0},
  doi       = {10.18653/v1/2025.acl-long.54},
  url       = {https://aclanthology.org/2025.acl-long.54/}
}

@inproceedings{paleyes2026code,
  title     = {Code Roulette: How Prompt Variability Affects
               {LLM} Code Generation},
  author    = {Paleyes, Andrei and
               Sendyka, Radzim and
               Robinson, Diana and
               Cabrera, Christian and
               Lawrence, Neil D.},
  booktitle = {Proceedings of the 3rd International Workshop on
               Large Language Models for Code},
  series    = {LLM4Code '26},
  year      = {2026},
  month     = apr,
  publisher = {Association for Computing Machinery},
  address   = {New York, NY, USA},
  location  = {Rio de Janeiro, Brazil},
  numpages  = {11},
  isbn      = {979-8-4007-2412-1},
  doi       = {10.1145/3786181.3788724},
  url       = {https://doi.org/10.1145/3786181.3788724}
}

@inproceedings{liu2023evalplus,
  author    = {Jiawei Liu and Chunqiu Steven Xia and Yuyao Wang and Lingming Zhang},
  title     = {Is Your Code Generated by {ChatGPT} Really Correct? Rigorous Evaluation of Large Language Models for Code Generation},
  booktitle = {Advances in Neural Information Processing Systems (NeurIPS)},
  volume    = {36},
  year      = {2023}
}

@article{chen2021humaneval,
  author  = {Mark Chen and Jerry Tworek and Heewoo Jun and Qiming Yuan and Henrique Ponde de Oliveira Pinto and Jared Kaplan and Harri Edwards and Yuri Burda and Nicholas Joseph and Greg Brockman and others},
  title   = {Evaluating Large Language Models Trained on Code},
  journal = {arXiv preprint arXiv:2107.03374},
  year    = {2021}
}

@article{austin2021mbpp,
  author  = {Jacob Austin and Augustus Odena and Maxwell Nye and Maarten Bosma and Henryk Michalewski and David Dohan and Ellen Jiang and Carrie Cai and Michael Terry and Quoc Le and Charles Sutton},
  title   = {Program Synthesis with Large Language Models},
  journal = {arXiv preprint arXiv:2108.07732},
  year    = {2021}
}

@article{ren2020codebleu,
  author  = {Shuo Ren and Daya Guo and Shuai Lu and Long Zhou and Shujie Liu and Duyu Tang and Neel Sundaresan and Ming Zhou and Ambrosio Blanco and Shuai Ma},
  title   = {{CodeBLEU}: A Method for Automatic Evaluation of Code Synthesis},
  journal = {arXiv preprint arXiv:2009.10297},
  year    = {2020}
}

@misc{haerpfer2022wvs,
  author       = {Christian Haerpfer and Ronald Inglehart and Alejandro Moreno and Christian Welzel and Kseniya Kizilova and Jaime Diez-Medrano and Marta Lagos and Pippa Norris and Eduard Ponarin and Bi Puranen},
  title        = {World Values Survey: Round Seven -- Country-Pooled Datafile},
  howpublished = {JD Systems Institute \& WVSA Secretariat, Madrid \& Vienna},
  year         = {2022},
  note         = {Version 4.0}
}

@article{dubey2024llama3,
  author  = {Abhimanyu Dubey and Abhinav Jauhri and Abhinav Pandey and Abhishek Kadian and Ahmad Al-Dahle and others},
  title   = {The Llama 3 Herd of Models},
  journal = {arXiv preprint arXiv:2407.21783},
  year    = {2024}
}

@article{jiang2023mistral,
  author  = {Albert Q. Jiang and Alexandre Sablayrolles and Arthur Mensch and Chris Bamford and Devendra Singh Chaplot and Diego de las Casas and others},
  title   = {Mistral {7B}},
  journal = {arXiv preprint arXiv:2310.06825},
  year    = {2023}
}

@article{openai2025gptoss,
  author  = {{OpenAI}},
  title   = {gpt-oss-120b \& gpt-oss-20b Model Card},
  journal = {arXiv preprint arXiv:2508.10925},
  year    = {2025}
}

@article{gemmateam2026gemma4,
  title         = {{Gemma 4 Technical Report}},
  author        = {{Gemma Team}},
  journal       = {arXiv preprint arXiv:2607.02770},
  year          = {2026},
  eprint        = {2607.02770},
  archivePrefix = {arXiv},
  primaryClass  = {cs.CL},
  url           = {https://arxiv.org/abs/2607.02770}
}

@misc{nvidia2025nemotron3nano30b,
  title        = {{NVIDIA-Nemotron-3-Nano-30B-A3B-BF16} Model Card},
  author       = {{NVIDIA}},
  year         = {2025},
  howpublished = {\url{https://huggingface.co/nvidia/NVIDIA-Nemotron-3-Nano-30B-A3B-BF16}},
  note         = {Accessed 2026-07-13}
}

@misc{nvidia2026nemotron3super120b,
  title        = {{NVIDIA-Nemotron-3-Super-120B-A12B-NVFP4} Model Card},
  author       = {{NVIDIA}},
  year         = {2026},
  howpublished = {\url{https://huggingface.co/nvidia/NVIDIA-Nemotron-3-Super-120B-A12B-NVFP4}},
  note         = {Accessed 2026-07-13}
}

\end{document}